# Growth of human population in Australia, 1000-10,000 years BP

Ron W Nielsen aka Jan Nurzynski[*]

Environmental Futures Centre, Gold Coast Campus, Griffith University, Qld, 4222, Australia

September 2013

Close analysis of the published interpretation of the number of rock-shelter sites in Australia provides further evidence that there was no intensification in the growth of human population between 1000 and 10,000 years BP. An alternative way of determining the time-dependent distribution of the size of human population between 1000 and 10,000 years BP is discussed.

In the earlier publication (Nielsen aka Nurzynski, 2013) we have discussed the common mistake, which is often made when interpreting hyperbolic-type distributions, the mistake consisting in seeing them as being made of two distinctly different trajectories, slow and fast, each requiring unnecessarily a different interpretation with an additional mandatory explanation of a non-existing transition between the two perceived components. We have

---

[*] r.nielsen@griffith.edu.au; ronwnielsen@gmail.com; http://home.iprimus.com.au/nielsens/ronnielsen.html

Suggested citation:

Nielsen, R. W. aka Nurzynski, J. (2013). *Growth of human population in Australia, 1000-10,000 years BP*. http://arxiv.org/ftp/arxiv/papers/1309/1309.0833.pdf

illustrated our discussion using the publication of Johnson and Brook (2011) who claimed the intensification in the growth of human population in Australia around 5000 years before present (BP). The discussion presented here contains a supplementary information not only about the claimed intensification but also about the alternative way of determining the size of human population in Australia between 1000 and 10,000 years BP.

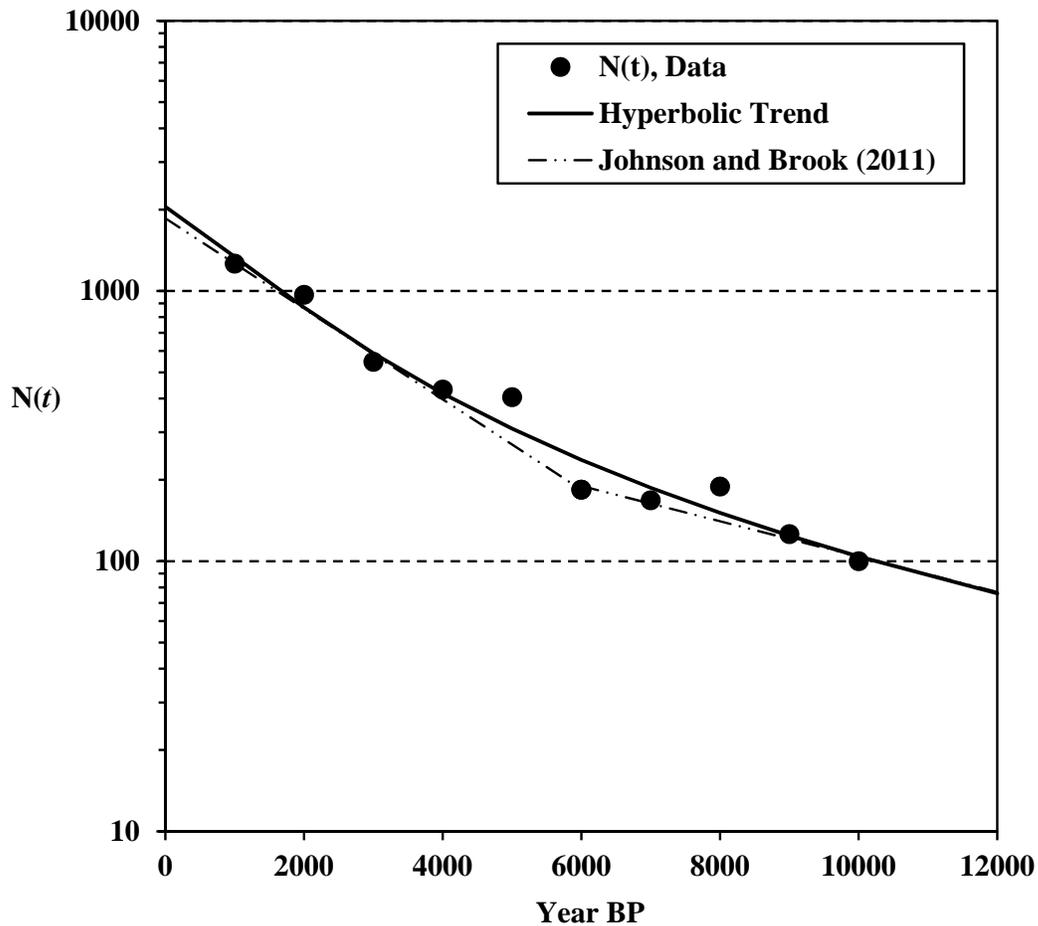

**Fig. 1.** The number of rock-shelter sites, $N(t)$, in Australia between 1000 and 10,000 years BP (Johnson & Brook, 2011) is compared with the best fit obtained using the 2nd-order hyperbolic distribution. The figure shows also the calculations of Johnson and Brook (2011, 2013) who claimed an abrupt change in the growth trajectory. The arrow of time is from right to left.

In Fig. 1 we are showing the time-dependent distribution of the number of rock-shelter sites in Australia, which were interpreted by Johnson and Brook (2011) as representing the size of human population. The figure compares two types of fits to the data: the fit using the 2nd-order hyperbolic distribution discussed earlier (Nielsen aka Nurzynski, 2013) and the best fit obtained by Johnson and Brook (2011, 2013). Their claim about the intensification of growth relies entirely on the high accuracy of just a single point at 6000 years BP. If this point is shifted up only a little, the claimed intensification disappears.

The best fit obtained by Johnson and Brook (2011) shown in Fig. 1 is represented by two exponential distributions joined at 6000 years BP. They have also tried to fit the data using two other mathematical descriptions:

$$N(t) = a + b \ln(t) \tag{1}$$

and

$$N(t) = a + \frac{b}{t} \tag{2}$$

where *a* and *b* are constants.

They call the eqn (1) "log-linear" and the eqn (2) "hyperbolic," which is however not a hyperbolic distribution but rather a *ratio* of two, very specifically chosen, hyperbolic distributions:

$$N(t) = a + \frac{b}{t} = \frac{N_1(t)}{N_2(t)} \tag{3}$$

where

$$N_1(t) = t^{-1} \tag{4}$$

and

$$N_2(t) = (at + b)^{-1} \quad (5)$$

None of the two equations [eqns (1) or (2)] fit the data (Johnson & Brook, 2013).

Our best fit, shown as a continuous line if Fig. 1, was obtained using the 2nd-order hyperbolic distribution

$$N(t) = (a_0 + at_1 + a_2t)^{-1} \quad (6)$$

where $a_0 = 0.000488$, $a_1 = 4.86 \times 10^{-7}$ and $a_2 = 7.255 \times 10^{-11}$.

The parameters used here are slightly different than in our earlier discussion (Nielsen aka Nurzynski, 2013) because previously we have used the data displaced by 500 years, exactly as published by Johnson and Brook (2011). In our present discussion we are using the data as supplied by Johnson and Brook (2013).

The data are of exceptionally good quality. This relatively high accuracy is more than adequate to study and determine the general trend, as obtained by fitting the 2nd-order hyperbolic distribution. However, to determine confidently a small deviation from this general trend and to claim the intensification of growth, one would need to have significantly higher quality data, arranged like beads along two distinctly different trajectories (one slow and one fast) or along a single trajectory, which would show a clear and unusual acceleration of growth at a certain time.

Close examination of the data presented in Fig. 1 indicated that the vertical deviations from the 2nd-order hyperbolic distribution, which can be considered as representing the general trend, can be as high as ±30%. The horizontal deviations are ±500 years because the data for the number of rock-shelter sites were binned in 1000-year intervals (Johnson and Brook, 2013). It is, therefore, unrealistic to assume a high accuracy for the single point at 6000 years BP and claim the intensification of growth around that time.

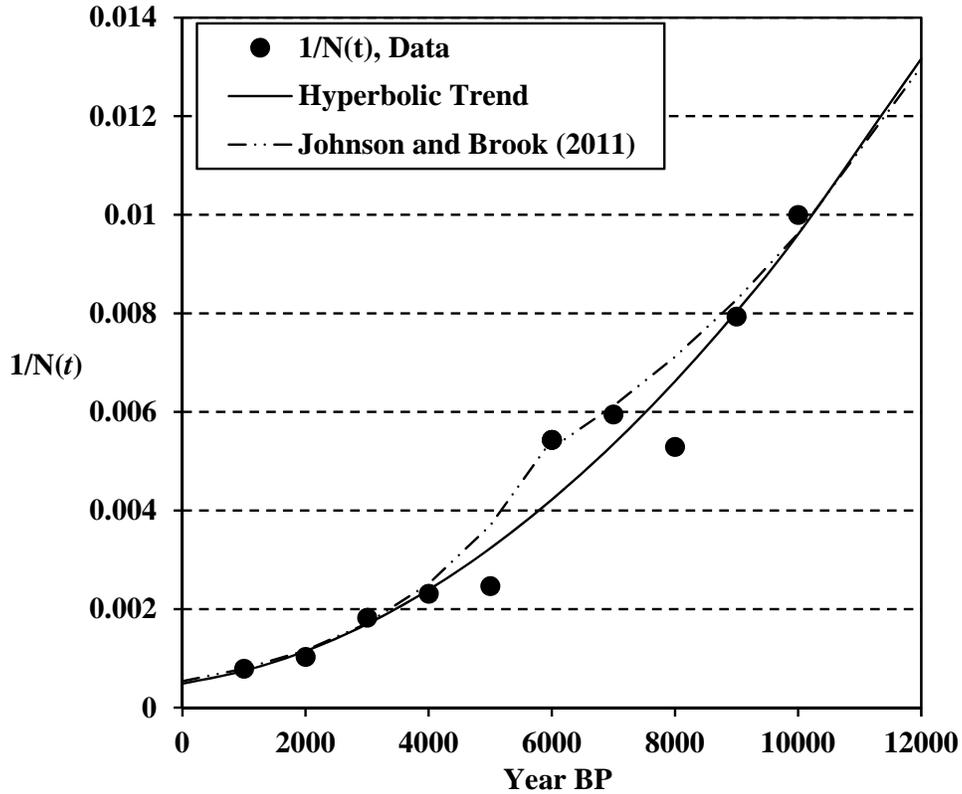

**Fig. 2.** The reciprocal values of the number of rock-shelter sites, $1/N(t)$, are compared with the reciprocal values of the 2nd-order hyperbolic distribution and the reciprocal values of the two trajectories calculated by Johnson and Brook (2011), all as displayed in Fig. 1.

The lack of support for the claimed intensification of growth can be demonstrated even more clearly by plotting the inverse values of the data and by comparing them with the relevant inverse values of the fitted distributions presented earlier in Fig. 1. Such a plot is shown in Fig. 2. The trajectories calculated by Johnson and Brook (2011) do not fit the data. There is also no obvious reason for selecting the point at 6000 years BP to serve as a linchpin for the two trajectories. The hypothesis of the intensification of growth of human population in Australia around 6000 years BP or at around any other time between 1000 and 10,000 years BP is not supported by data.

We shall now focus on the estimation of the growth of human population. Johnson and Brook (2011) assumed that the growth of human population was represented directly by the number of rock-shelter sites. The larger is the number of rock-shelter sites the larger is the size of the population. It is a reasonable assumption and in its general form it can be represented by the linear relationship

$$S(t) = aN(t) + b \qquad (7)$$

where $S(t)$ is the size of human population, $N(t)$ is the number of rock-shelter sites, and $a$ and $b$ are constants.

Johnson and Brook (2011) assumed implicitly that $a = 1$ and $b = 0$. This assumption leads to a time-dependent distribution of the size of human population, which can be described well by the 2nd-order hyperbolic distribution (see Fig. 1).

The parameter $a_2$ in the eqn (6) is small but it is still not equal zero. It is unclear why the 2nd-order hyperbolic distribution should be used to describe the growth of human population in Australia. A simpler description would be more appealing. The small value for the parameter $a_2$ suggests that perhaps with a slightly modified way of relating the number of rock shelters to the size of the population one could use a simpler formula.

Indeed, it turns out that only a minor modification is required. All we have to do is to use $b = 110$ in the eqn (7) while keeping the parameter $a$ at $a = 1$. The relation between the number of rock shelters and the size of human population is now given by

$$S(t) = N(t) + 110 \qquad (8)$$

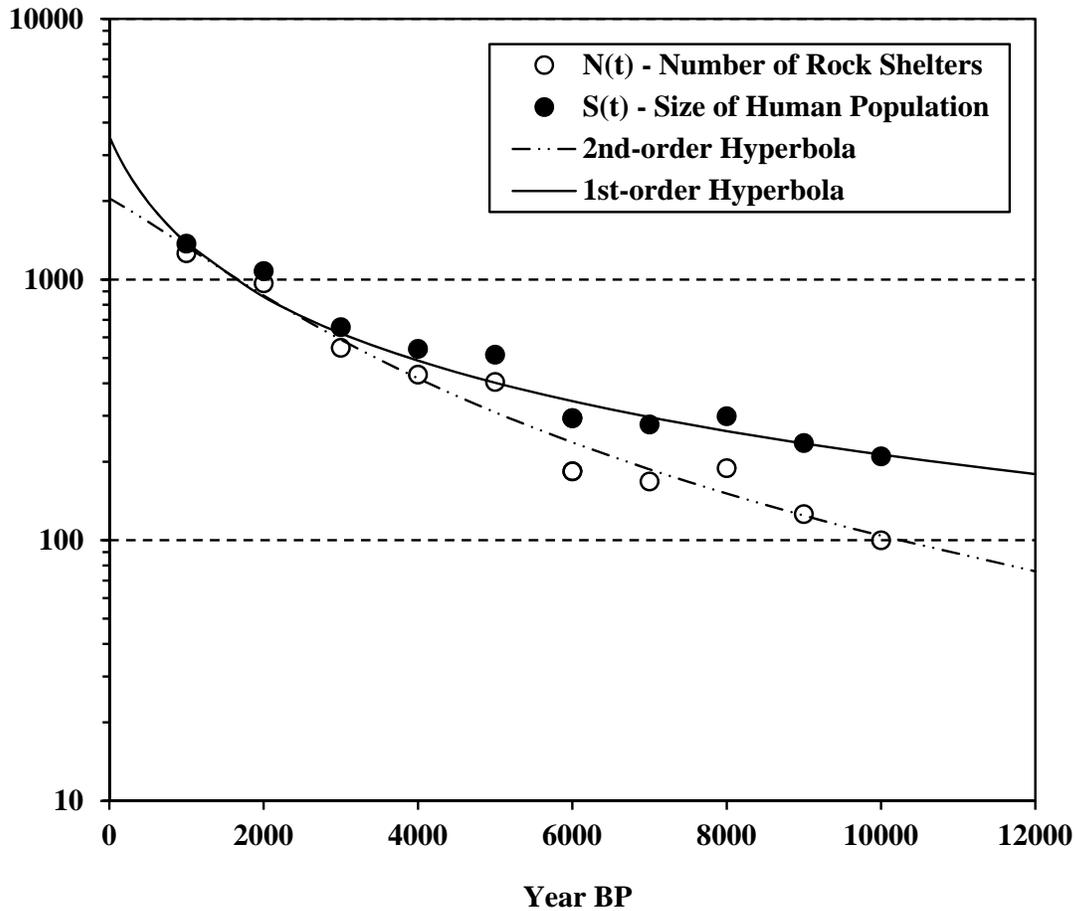

**Fig. 3.** The number of rock-shelter sites, $N(t)$, (Johnson & Brook, 2011) and the corrected size of human population, $S(t)$, determined using the eqn (8) are compared with hyperbolic distributions.

The eqn (8) represents a purely empirical formula but it has a simple interpretation. It suggests that a systematic error is probably made when relating the size of the population to the number of rock shelters. If we apply this correction then the growth of human population can be described by a simpler, 1st-order hyperbolic distribution. Two distributions, one representing the number of rock shelters or the uncorrected size of human population and one representing the corrected size are shown in Fig. 3. The 1st-order hyperbolic distribution fitting the population data is

$$S(t) = (a_0 + a_1 t)^{-1} \tag{9}$$

where $a_0 = 0.000282$ and $a_1 = 4.4134 \times 10^{-7}$.

In summary, this study shows conclusively that there was no intensification in the growth of human population in Australia around 6000 years BP or at any other time between 1000 and 10,000 years BP. It also shows that with only one minor modification to the calculations of the size of human population from the number of rock-shelter sites, the growth of human population in Australia between 1000 and 10,000 years BP can be described well using a simple, 1st-order hyperbolic distribution.

No claim is made that the "corrected" distribution is more accurate than the "uncorrected" distribution. The only feature, which is more appealing is that the "corrected" distribution can be described using a simpler mathematical formula. To distinguish between the two distributions one would have find an independent way of estimating the size of the population around 10,000 BP where the difference between the two distributions is large.

Helpful correspondence with Chris Johnson and Barry Brook, who supplied the data for the number of rock-shelter sites and additional information about their calculations, is gratefully appreciated.